\documentstyle[aps,twocolumn,epsfig]{revtex}
\begin{document}
\newcommand{\ve}[1]{\mbox{\boldmath $#1$}}
\twocolumn[\hsize\textwidth\columnwidth\hsize
\csname@twocolumnfalse%
\endcsname

\draft

\title{Propagation of exciton pulses in semiconductors}
\author{A. D. Jackson$^1$ and G. M. Kavoulakis$^{2,*}$}
\date{\today}
\address{$^1$Niels Bohr Institute, Blegdamsvej 17, DK-2100 Copenhagen \O,
        Denmark, \\
         $^{2}$Royal Institute of Technology, Lindstedtsv\"agen 24,
         S-10044 Stockholm, Sweden}

\maketitle
			  
\begin{abstract}

Using a toy model, we examine the propagation of excitons in Cu$_2$O, which 
form localized pulses under certain experimental conditions.  The
formation of these waves is attributed to the effect of dispersion,
non-linearity and the coupling of the excitons to phonons, which
acts as a dissipative mechanism.

\end{abstract}

\pacs{PACS numbers: 05.45.Y, 05.30.Jp, 71.35.-y}

\vskip1.5pc]

\section{Introduction}

The transport properties of excitons \cite{Knox} in Cu$_2$O
have been studied extensively in recent years. In one series
of experiments \cite{Fortin,Andre,Benson,Benson2,Alex}, a pulse
of excitons was created at one end of a crystal of Cu$_2$O
and detected at the other following propagation inside the sample.
In another series of experiments \cite{TWM,TW,SWM}, excitons were
created and imaged as they expanded inside a crystal subjected to
external stress.  (The stress effectively created a confining
potential for the excitons.)

Two noteworthy conclusions emerged from these experiments.  First,
drag in the propagation of excitons due to their scattering with
(longitudinal) acoustic phonons was found to decrease with decreasing
lattice temperature.  The second observation \cite{Andre,Benson,Benson2}
followed from experiments with two laser pulses separated by a time
delay of $\sim 100$ ns. The first pulse created an asymmetric wave of
excitons; the excitons created by the second pulse travelled
in the trail of the first.  This second wave was observed to be symmetric
and to have the characteristics of a solitary-wave, specifically, those
of a ``bright soliton" (i.e., a local elevation in the density.)

Here, we will argue that the waves created by the second pulse are likely
to resemble classical solitary waves.  These waves are, however, essentially
different from solitary waves in, e.g., water (as first observed by J.
Scott Russell \cite{SR}) due to the presence of dissipation caused by the
scattering of excitons with phonons \cite{TWM,TW,SWM}.  We will suggest that
the effects of nonlinearity and dispersion, which give rise to solitons,
must be combined with dissipation in order to create generic conditions
appropriate for the propagation of localized pulses of excitons in the
crystal.  All three mechanisms play an important role.

The studies of Refs.\,\cite{EH,LR} attributed the observed localized waves
to the formation of a Bose-Einstein condensate of excitons \cite{MS}.  In
Ref.\,\cite{EH}, second-order perturbation theory was used to describe
the exciton-exciton interaction \cite{EH}, which gave rise to effective
attraction between the condensed excitons. In Ref.\,\cite{LR}, it was shown
that an exciton condensate coupled to acoustic phonons can give rise to
bright solitons.  However, as pointed out in Ref.\,\cite{T}, solitary waves
in a Bose-Einstein condensed gas have a characteristic length scale
determined by its healing length $\xi$ given by $\hbar^2/2 m \xi^2=n U_0$,
or $\xi = (8 \pi n a)^{-1/2}$, where $m$ is the exciton mass, $n$ is the
exciton condensate density, and $U_0=4 \pi \hbar^2 a/m$ is the exciton-exciton
scattering matrix element, with $a$ being the scattering length.  For typical
liquid helium temperatures of order 2 K, the critical density for the
Bose-Einstein condensation of excitons is of order $10^{17}$ cm$^{-3}$. The
scattering length for exciton-exciton elastic collisions in Cu$_2$O has been
calculated to be $a \approx 2 a_B$~\cite{SC}, where $a_B$ is the exciton Bohr
radius.  The Bohr radius of excitons in Cu$_2$O is $\approx 5$
\AA\,\cite{KCB} so that $\xi \sim 0.01$ $\mu$m, which is about four orders
of magnitude smaller than the width of the narrowest pulses $l_p \approx
1$ mm.  (The crystal is typically a few millimeters long.)

Finally, a model which has been proposed in connection with the ballistic
propagation of excitons attributes the observed phenomenon to a classical
effect due to a ``phonon wind"~\cite{BT}. Many of the observations have
been explained in the context of this model and our study is consistent
with the claim that the excitons are described by classical equations.
On the other hand, the two studies examine effects which are essentially 
decoupled.

More specifically, while both models include dissipation, the phonon-wind
model focuses on the spatial and temporal effects of laser excitation,
whereas our model concentrates on the effects of non-linearity and dispersion.
In the present study we make the implicit simplified assumption 
of a delta-like form of laser excitation in space and time. In such a case,
the phonon-wind model essentially reduces to a diffusion-like equation, 
neglecting the effects of non-linearity and dispersion. Of course, one should
worry about more complicated forms of the laser excitation, however the 
problem we have studied is quite complicated and we have chosen 
to overlook these effects in order to demonstrate clearly the combined
effect of dissipation, non-linearity and dispersion; we believe all three 
effects are very crucial in these experiments. Finally, from an 
experimental point of view, our study would be mostly suitable for 
short (in space and time) laser pulses.

\section{Hydrodynamic description}
Following Ref.\,\cite{LB}, we start with the hydrodynamic equations
describing the system,
\begin{eqnarray}
  \frac {\partial n}  {\partial t} + {\bf \nabla} \cdot (n {\bf v}) &=& 0,
\nonumber \\
 \left[ \frac {\partial} {\partial t} + {\bf v} \cdot {\bf \nabla}
 \right] {\bf v} &=& - \frac {{\bf \nabla} p} n - \frac {{\bf v}} {\tau},
\label{hy}
\end{eqnarray}
where ${\bf v}$ is the velocity field, $p$ is the pressure, and
$\tau$ is the exciton-acoustic phonon scattering time. The neglect of
viscosity in these equations can be justified by looking at
the characteristic scales: The viscous force $\eta \nabla^2 v$
is of order $\eta v / l_p^2 \sim n \tau_s c^3 / l_p^2$, where $\eta$ is
the viscosity, $\tau_s$ is the scattering time for elastic exciton-exciton
collisions, and $c^2=\partial p/ \partial n$ is the adiabatic sound velocity.
Therefore, $|\eta \nabla^2 v / \nabla p| \sim c \tau_s / l_p$.  Typical
values of $c\sim 5\times10^5$ cm/s, $l_p \sim$ 1 mm, and $\tau_s=
(n \sigma v)^{-1}$ with $\sigma = 8 \pi a^2$ imply that $\tau_s \sim$ 1 ns
for $n \sim 10^{17}$ cm$^{-3}$, $a\sim10$ \AA, and a temperature $T \sim 1$
K. Therefore, $c \tau_s / l_p \sim 10^{-2} \ll 1$.

Linearizing the above equations and considering small variations in the
density and the pressure, $n= n_0 + \delta n$ and $p= p_0 + \delta p$
(where $n_0$ and $p_0$ are the ``background"  density and pressure
of excitons created by the first laser pulse), one can derive the equation:
\begin{eqnarray}
    \frac {\partial^2 \delta n} {\partial t^2} - c^2 {\bf \nabla}^2
 \delta n + \frac 1 {\tau} \frac {\partial \delta n} {\partial t} = 0.
\label{lin}
\end{eqnarray}
This equation leads to the dispersion relation
\cite{LB}
\begin{eqnarray}
  \omega^2 - c^2 k^2 + i \omega / \tau = 0,
\label{dis}
\end{eqnarray}
which interpolates between two limits.  For $k c \tau \ll 1$,
the motion of the exciton pulse is diffusive, and
\begin{eqnarray}
  \partial_t {\delta n} = c^2 \tau {\bf \nabla}^2 \delta n,
\label{dif}
\end{eqnarray}
with $\omega = - i c^2 \tau k^2$.
In the opposite limit $k c \tau \gg 1$,
\begin{eqnarray}
 {\partial_{tt}} {\delta n}  - c^2 {\bf \nabla}^2
 \delta n = 0,
\label{wav}
\end{eqnarray}
with $\omega = c k$, corresponding to sound propagation.

\section{Non-linearity versus dispersion}  
The analysis so far applies to
small variations around equilibrium.  For larger excitations it is also
necessary to account for the effects of non-linearity and dispersion.
For simplicity, we start with a (one-dimensional) wave propagating
with velocity $c$ along the positive $x$ axis,
\begin{eqnarray}
  \partial_t \delta n +
 c \, \partial_x \delta n = 0.
\label{wave}
\end{eqnarray}
Any function of the form $\delta n = \delta n(x - ct)$ is a solution
of Eq.\,(\ref{wave}).  Allowing for dispersion and non-linearity, this
becomes
\begin{eqnarray}
   \frac{\partial \delta n}{\partial t} +
   c (1 + \alpha \frac{\delta n}{n_0}) \frac{\partial \delta n}{\partial x}
   + \beta \frac{l_p^3}{\tau_p} \frac{\partial^3  \delta n}{\partial x^3} = 0.
\label{wavend}
\end{eqnarray}
Here $\alpha$ and $\beta$ are small (positive) dimensionless quantities,
which can be calculated from microscopic theories. It is not our goal to 
calculate them here, and we treat them as phenomenological parameters. Also
$\tau_p$ is the time for the pulse to travel a length $l_p$ (i.e.,
$l_p = c \tau_p$.)  Typically, $\tau_p \sim 100$ ns for relatively narrow
pulses. Equation (\ref{wavend}) leads to the dispersion relation
\begin{eqnarray}
   \omega = c k \left[ 1 + \alpha \frac {\delta n} {n_0}
 - \beta (k l_p)^2 \right],
\label{dnl}
\end{eqnarray}
which provides a reasonable description of non-linearity and dispersion.
For $k \to 0$, the sound velocity is $c(1+\alpha \delta n/n_0)$, and
the local sound velocity is greater in regions of higher density.  For
$\alpha=0$, $\omega / k = c [1-\beta (k l_p)^2]$, which implies that
longer wavelengths travel faster.  Equation (\ref{wavend}) is of
standard Korteweg--de Vries form, and the competition between the last
two terms gives rise to solitonic solutions \cite{sol}.

We seek solutions of Eq.\,(\ref{wavend}) which preserve their
shape, i.e., which have the form $\delta n(x-ut)$, and which satisfy
the boundary condition that $\delta n \to 0$ as $|z| \to \infty$.
Such solutions have the form
\begin{eqnarray}
   \frac {\delta n} {n_0} = C\,
  {\rm sech}^2 \left[ \left( \frac {\alpha C} {12 \beta}
\right)^{1/2} \frac {x - u t} {l_p} \right],
\label{newequuint2}
\end{eqnarray}
with $u=c(1+\alpha C /3)$.  For fixed $\alpha$ and $\beta$, there is thus
a family of solutions characterized by their maximum amplitude, $C$.  For
$\alpha C \sim 12 \beta$ the size of the travelling wave is of order
$l_p$.  This result reflects the expected competition between non-linearity
and dispersion that gives rise to this localized excitation. In addition, the
velocity of this pulse is of order $c$, which can approach the speed of sound
$c_l$ of Cu$_2$O.  For example, for a classical gas $p = n k_B T$, and thus
$c^2 = k_B T / m$, with $m= 3 m_e$ being the total exciton mass density
($m_e$ is the electron mass.)  For $T \approx 2$ K, $c \approx 3.2 \times
10^5$ cm/s, in good agreement with experiment.

\section{Scattering of excitons with the crystal}
Equation (\ref{wavend}) does not include the effects of the scattering of
excitons with the crystal.  As noted above, this mechanism leads to
the diffusive motion of excitons.  It is thus worth considering what
is known about exciton-phonon interactions in Cu$_2$O.

In Refs.\,\cite{TWM,TW}, paraexcitons under low excitation conditions were
observed to be highly mobile, with a diffusion constant $D \sim 10^3$ cm$^2$/s.
For $D=c^2 \tau$ with $c^2 = k_B T/m$, $\tau$ is seen to be
$\approx 20$ ns. For typical thermal velocities of order $3 \times 10^5$
cm/s, the mean-free path of excitons is $\approx 50$ $\mu$m.  This
corresponds to $\approx 10^5$ lattice sites and is thus quite large.
In addition, $D$ showed a rapid increase for temperatures below
$\approx 6$ K.  These observations were explained in Ref.\,\cite{TW}
using a simple classical model which showed that excitons with
wavevectors smaller than $k_l=m c_l/ \hbar$, where $c_l$ is the
longitudinal speed of sound in Cu$_2$O ($c_l \approx 4.5 \times 10^5$
cm/s), cannot emit acoustic phonons and thus cannot scatter with the
lattice.  For exciton temperatures below $T_l$, where $k_B T_l \approx
m c_l^2$, exciton-acoustic phonon scattering is suppressed.  This
freeze-out temperature is $T_l \approx 4$\,K for longitudinal acoustic
phonons in Cu$_2$O.

In the experiments of Refs.\,\cite{Fortin,Andre,Benson,Benson2}
with an excitation of $\approx 1.2 \times 10^6$ W/cm$^2$, the temperature
varied between 1.9 K and 4.1 K. For $T \approx 2.5$ K, which is close to
the freeze-out temperature $T_l$, the propagation of the exciton pulse
showed a transition from diffusive to ballistic motion. In Ref.\,\cite{Benson}
for conditions under which the exciton propagation was diffusive, the
diffusion constant was reported to be even higher, $D \approx 2 \times
10^3$ cm$^2$/s, corresponding to $\tau \approx 40$ ns.

It is worth pointing out that in the experiments where the sample was
illuminated with two pulses \cite{Andre,Benson,Benson2}, these were
delayed by $\sim 100$ ns. This time delay is comparable to $\tau_p$.
It is thus reasonable to expect that the first pulse diffuses inside
the crystal and creates a background density of excitons through which
the second pulse must pass.

\section{A more realistic problem}
In view of these considerations, we have included an additional term
in Eq.\,(\ref{wavend}) to describe the effects of exciton-phonon
scattering:
\begin{eqnarray}
   \frac {\partial \delta n} {\partial t} +
  c (1 + \alpha \frac {\delta n} {n_0}) \frac {\partial \delta n} {\partial x}
 + \beta \frac {l_p^3} {\tau_p} \frac {\partial^3  \delta n} {\partial x^3} =
 \gamma \frac {l_p^2} {\tau_p} \frac {\partial^2  \delta n} {\partial x^2},
\label{wavenddis}
\end{eqnarray}
with $\gamma$ being a positive parameter.  Here $\tau_p/\gamma$ can be
identified as the exciton-acoustic phonon scattering time.  Since
$\tau_p \sim 100$ ns and $\tau$ is $\sim 30$ ns, $\gamma$ is expected
to be greater than, but on the order of unity.  The term on the right side 
of Eq.\,(\ref{wavenddis}) gives an imaginary part to the dispersion relation, 
modifying Eq.\,(\ref{dnl}) to
\begin{eqnarray}
   \omega = c k \left[ 1 + \alpha \frac {\delta n} {n_0}
 - \beta (k l_p)^2 \right] + (k l_p)^2 \frac {i \gamma} {\tau_p}.
\label{dnll}
\end{eqnarray}
With such a dispersion relation the wave is damped. For example for
$k \sim 1/ l_p$, $\delta n \sim e^{-\gamma t/ \tau_p}$.

\section{Effect of dissipation on the pulse propagation}  
Equation (\ref{wavenddis}) interpolates between two extreme modes of
propagation.  For $\gamma=0$ the solution is a travelling wave, which
propagates without change in shape with a velocity of order $c$.  When
$\gamma$ is nonzero, the width of the pulse increases with time and
its shape changes.  As $\gamma$ is increased, this spreading velocity
increases and eventually equals $c$ at some $\gamma_{\rm max}$.  For
$\gamma \ll \gamma_{\rm max}$, the pulse pulse spreads slowly relative
to its propagation velocity of $\approx\,c$.

To determine $\gamma_{\rm max}$, it is convenient to use the notation
$\phi= \delta n/n_0$ and the quantities
\begin{equation}
\langle X \rangle = \frac{\int x \phi \, dx}{\int \phi \, dx} \ \ \ {\rm and}
\ \ \ \langle X^2 \rangle = \frac{\int x^2 \phi \, dx}{\int \phi \, dx},
\label{5}
\end{equation}
and the associated width $\Delta X = \sqrt{\langle X^2 \rangle -\langle X
\rangle^2}$.  The time derivative $\partial_t \langle X \rangle$ can be
regarded as the velocity of propagation of the pulse; $\partial_t \Delta X$
describes the rate of increase of the width.  The desired average values
follow immediately from appropriate integrals of Eq.\,(\ref{wavenddis}).
We find that
\begin{equation}
\frac{\partial}{\partial t} \langle X \rangle = c \left( 1 + \frac {\alpha} 2
\frac{\int \phi^2 \, dx}{\int \phi \, dx} \right).
\label{7}
\end{equation}
For $\gamma = 0$ and $\phi$ given by Eq.\,(\ref{newequuint2}),
Eq.\,(\ref{7}) gives the expected result that $\partial_t \langle X \rangle
= c (1 + \alpha C /3)$, i.e., the velocity of propagation $u$ given above.
Similarly, we find that
\begin{equation}
   \frac{\partial}{\partial t} \Delta X^2 = \alpha c \left(
  \frac{\int \, x \phi^2 \, dx}{\int \, \phi \, dx} -
 \frac{\int \, x \phi \, dx}{\int \, \phi \, dx}
\frac{\int \, \phi^2 \, dx}{\int \, \phi \, dx} \right)
 + 2 \gamma c l_p.
\label{3}
\end{equation}
The $\alpha$-dependent term in this equation evidently vanishes for any
symmetric pulse shape.  For such pulses we see that
\begin{equation}
\partial_t \Delta X = \frac{\gamma c l_p}{\Delta X}.
\label{3a}
\end{equation}
For pulses of typical width of order $l_p$, this suggests that
$\gamma_{\rm max}$ is of order $1$.  The characteristic scattering time
of $\approx\, \tau_p$, which establishes the crossover from the motion of
a localized pulse to diffusive motion, is thus of the same order as the
characteristic scattering time $\tau$ for exciton-phonon collisions that
has been determined experimentally as described above.  As we discuss
below, the approximately linear growth of $\Delta X^2 (t)$ indicated by
Eq.\,(\ref{3}) provides a definite test for the validity of our
description.

\section{The benefits of dissipation}  
While dissipation can obscure the
solitonic propagation, it also plays a far more positive role.  In its
absence, an initially localized pulse will often break up into a discrete
number of solitons (plus small amplitude, non-solitonic motion.)  This is
illustrated in Fig.\,1(a) where we show the evolution of an initial pulse
\begin{equation}
\phi (x,0) = \frac{\lambda}{3}\exp{\left( - \frac{x^2}{\lambda^2 l_p^2}
\right) }
\label{pulse}
\end{equation}
with $\lambda = 3$, which leads to the formation of three solitons.
(The calculation was performed for $\alpha = 0.24$, $\beta = 0.02$, and
$\gamma = 0$.)  Once formed, the distance between
the individual solitons grows linearly with
$t$.  The generic break-up of localized peaks into many solitons is in
marked contrast to the experiments considered here, which see only single,
localized peaks.  Dissipation of sufficient strength can provide an effective
mechanism for preventing the formation of unwanted local maxima.  This is
illustrated in Figs.\,1(b)--(d) obtained numerically for $\gamma =
0.01$, $0.03$, and $0.2$,  respectively.  Once new local extrema have
been formed, dissipation will tend to reduce the local maximum (where the
curvature is negative) and enhance the local minimum (where the curvature is
positive.)  For all values of $\gamma$, the local maximum and minimum slowly
merge and ultimately disappear.  Given the linear growth of the separation
between local maxima and the much longer time scale required
for dissipation to eliminate them, it is clear that $\gamma$ should
exceed some minimum value, $\gamma_{\rm min}$, chosen to ensure
that secondary maxima never form.  The value of $\gamma_{\rm min}$ is set
by a complicated interplay between the initial pulse shape and the parameters
appearing in Eq.\,(\ref{wavenddis}).  For the present case of $\lambda = 3$,
numerical simulations indicate that $\gamma_{\rm min} = 0.024$.  The initial
pulse width considered here is larger than that appropriate to the experimental
initial conditions of Refs.\,\cite{Fortin,Andre,Benson,Benson2}.
A more realistic calculation would involve a value of $\lambda$ that is
smaller than (but on the order of) unity.
We have used here the value $\lambda = 3$ for convenience.
In addition, numerical evidence indicates that $\gamma_{\rm min}$ grows 
{\it slowly} as the width $\lambda$ decreases.

An interesting feature of our results is that $\gamma_{\rm max}$ and
$\gamma_{\rm min}$ differ by a large factor (i.e., $40$ in the present case.)
This suggests that it is not necessary to fine tune the dissipative term in
Eq.\,(\ref{wavenddis}) in order to obtain localized pulse propagation.
Indeed, the qualitative impression of solitary pulse propagation seems to
be a particularly robust phenomenon, and its observation cannot be regarded
as a demonstration of true solitonic propagation.

\section{Two experimental challenges}
The model we have presented can
be checked easily since it makes the following two definite
predictions: 1) We have noted that, in the presence of
dissipation, $\Delta X^2(t)$ grows linearly with $t$.  Since $t$ is the
time for the pulse to propagate inside the crystal, $\Delta X^2$ should
also grow linearly with the size $L$ of the crystal along the direction
of pulse propagation (keeping all other conditions fixed.)  2) As we
have seen, the propagation of excitons is sensitive to the
value of $\gamma$, and thus to their coupling with phonons.  If an
external stress (of a few kbar) is applied, excitons can couple to
transverse acoustic phonons, as discussed in Ref.\,\cite{TW}.  The transverse
speed of sound $c_t$ is 
\noindent
\begin{figure}
\begin{center}
\epsfig{file=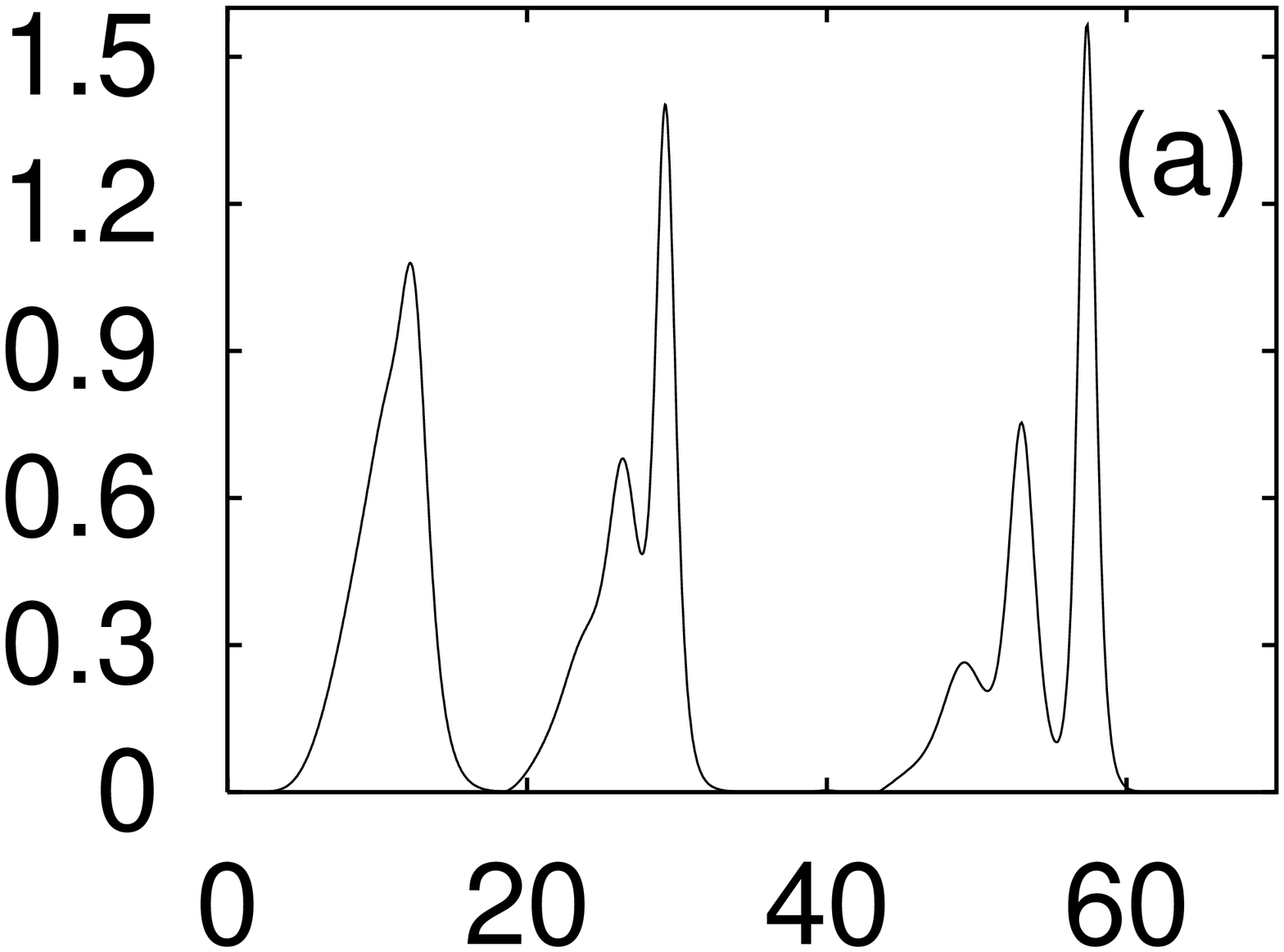,width=4.0cm,height=8.0cm,angle=-90}
\epsfig{file=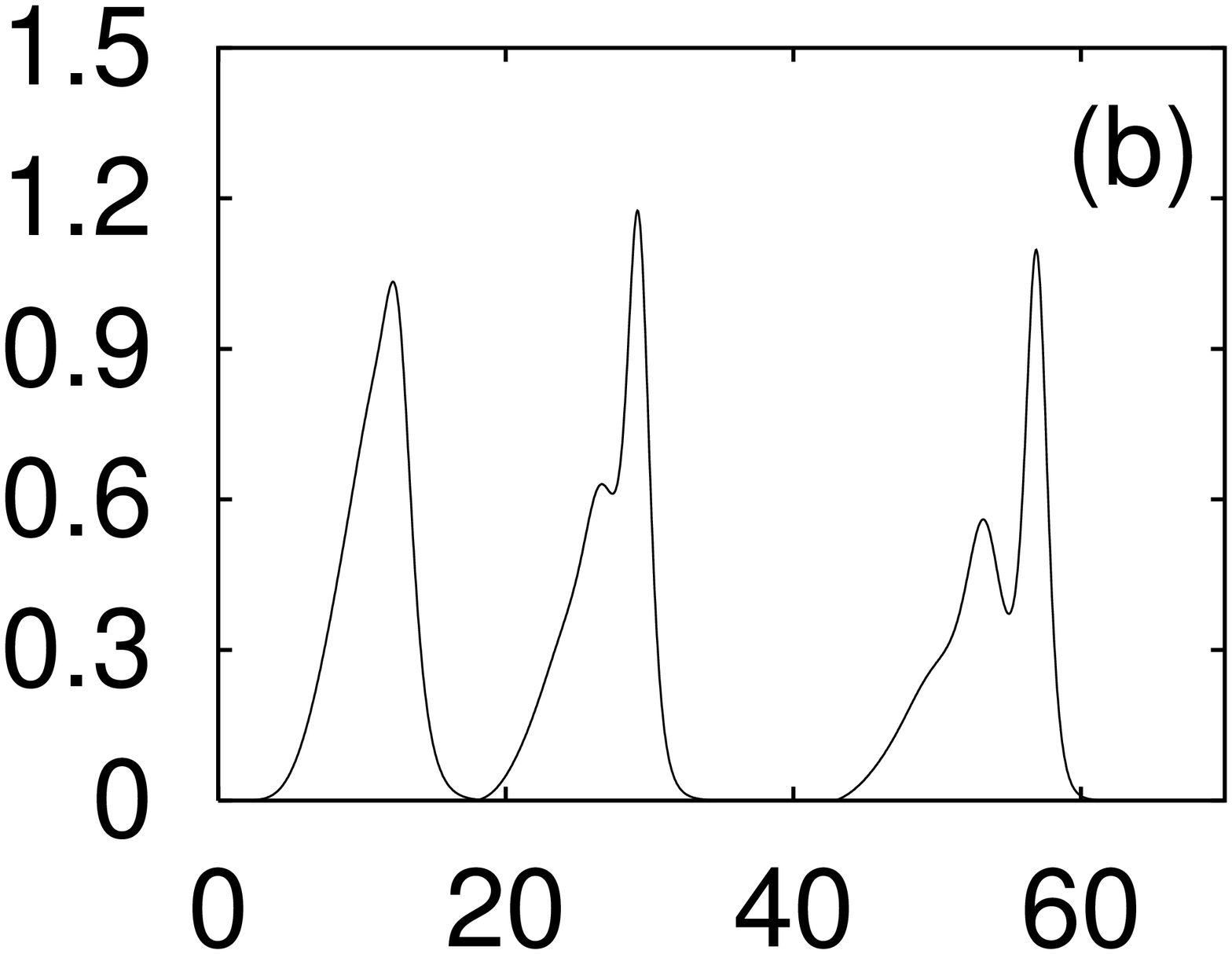,width=4.0cm,height=8.0cm,angle=-90}
\epsfig{file=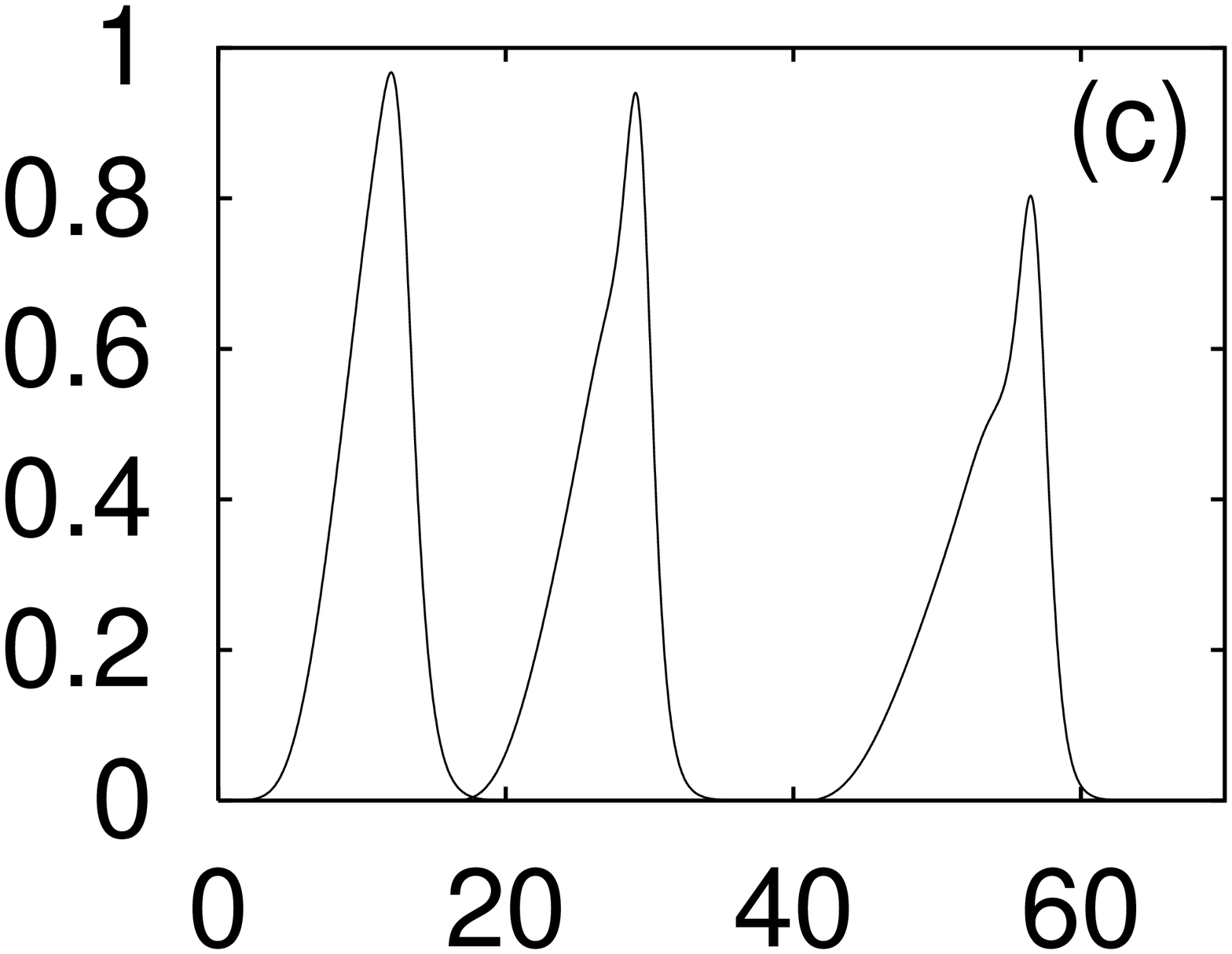,width=4.0cm,height=8.0cm,angle=-90}
\epsfig{file=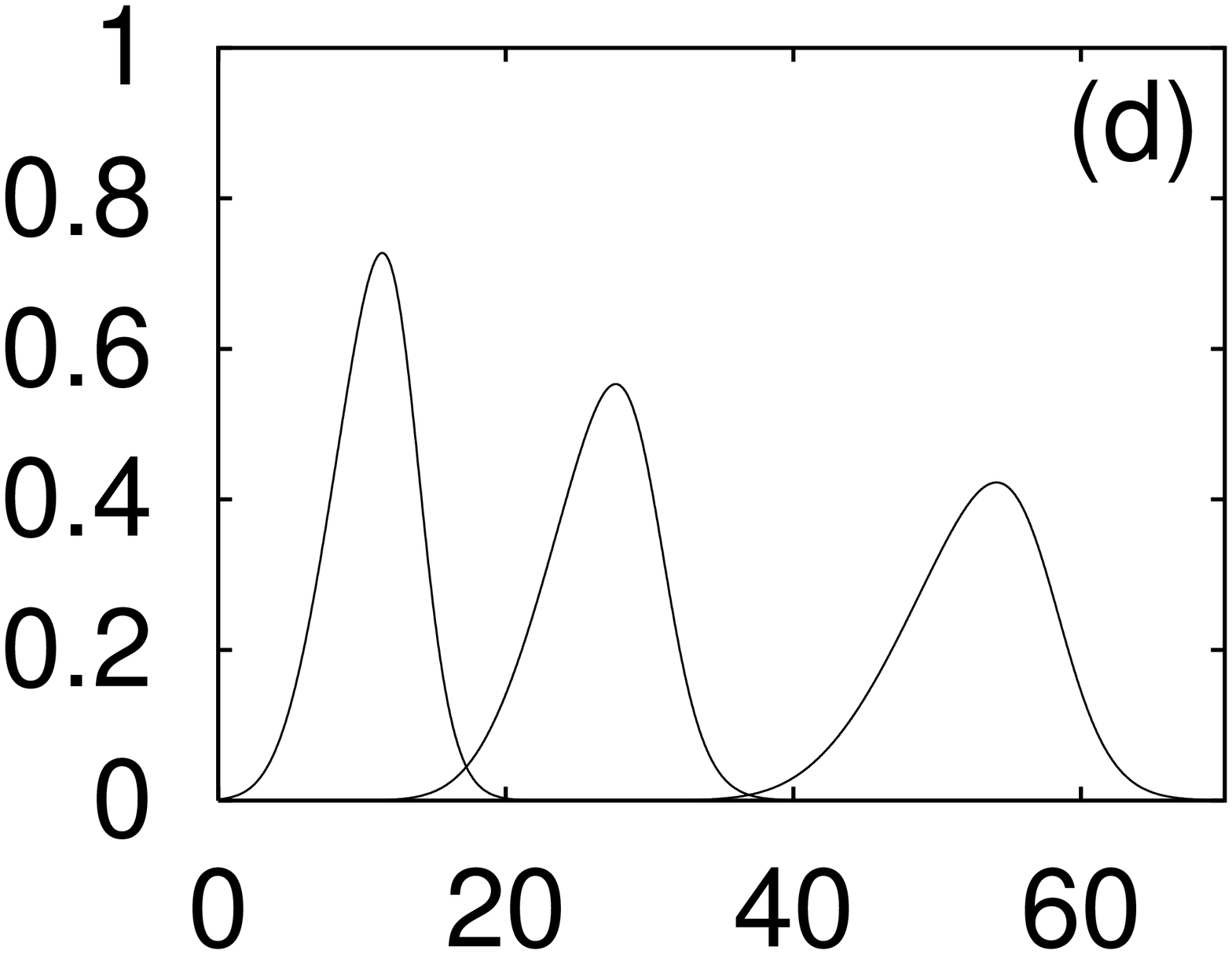,width=4.0cm,height=8.0cm,angle=-90}
\vskip0.5pc
\begin{caption}
{Solutions of Eq.\,(\ref{wavenddis}). We have plotted $\delta n/ n_0$
($y$ axis) as function of distance ($x$ axis), measured in units of
$l_p$. The initial disturbance is given by Eq.\,(\ref{pulse}) with $\lambda
= 3$. We have used the values $\alpha = 0.24$ and $\beta=0.02$. The figures
show snapshots of the solution for $\gamma = 0$ (a), $0.01$ (b), $0.03$ (c),
and $0.2$ (d), at times $t/\tau_p = 10.0,\ 25.0$, and $50.0$.}
\end{caption}
\end{center}
\label{FIG1}
\end{figure}
\noindent
lower than $c_l$, and the corresponding
freeze-out temperature is consequently some ten times smaller than
$T_l$, i.e., $\approx 0.5$ K \cite{TW}.  If the same experiments were
performed in the presence of stress, the term on the right of
Eq.\,(\ref{wavenddis}) would be enhanced and the motion would become
more diffusive with increasing stress.  (The conclusions of the other
theories \cite{EH,LR,BT} are expected to be less sensitive to external
stress.)  These tests are likely to provide compelling evidence for or
against our picture since the empirical value of $\gamma$, estimated above,
is relatively close to $\gamma_{\rm max}$.

\section{Summary} We have presented a phenomenological theory attributing the
formation of localized waves of excitons in Cu$_2$O under certain
experimental conditions to the combined effects of dispersion,
non-linearity and dissipation.  Both analytical estimates and
numerical simulations suggest that the characteristic time scale for
the crossover between the regimes of localized propagation of the
pulse and diffusive motion is comparable to the scattering time for
exciton-phonon collisions.

\acknowledgments
G.M.K. is grateful to G. Baym, A. Jolk, M. J\"orger, and
C. Klingshirn for useful discussions. G.M.K. would also like
to thank the Physics Department of the Univ. of Crete, Greece
for its hospitality.

\vskip1pc
\noindent
$^*$ Present address: Mathematical Physics, Lund Institute of Technology, P.O.
Box 118, S-22100 Lund, Sweden

\end{document}